\documentclass[prd,aps,amsmath,amssymb,twocolumn,groupedaddress,floats,showpacs,final]{revtex4-1}
\usepackage{graphicx}
\usepackage{mathrsfs}
\usepackage{subfigure}
\usepackage{dcolumn}
\usepackage{amsmath}
\usepackage{tabularx}
\usepackage{bm}
\usepackage{color}
\usepackage{hyperref}
\usepackage{cleveref}

\newcommand{\vk}{{\mathbf{k}}}

\begin{document}

\author{Kun Chen}
 \email{chenkun@itp.ac.cn}
 \affiliation{CAS Key Laboratory of Theoretical Physics, Institute of Theoretical Physics, Chinese Academy of Sciences, Beijing 100190, China}
  \affiliation{Department of Physics and Astronomy, Rutgers, The State University of New Jersey, Piscataway, NJ 08854-8019 USA}

\title{Partial Renormalization of Quasiparticle Interactions}

\date{\today}

\begin{abstract}

Nonlocal effective interactions are inherent to non-relativistic quantum many-body systems, but their systematic resummation poses a significant challenge known as the ``vertex problem" in many-body perturbation theory. We introduce a renormalization scheme based on a projection-based renormalization condition that selectively resums the most essential nonlocal contributions to the effective interaction vertex, avoiding the computational complexity of the full vertex function. This enables us to derive a renormalized Feynman diagrammatic series with large parameters canceled by counter-diagrams, efficiently generated using a perturbative expansion of the parquet equations and computed using a diagrammatic Monte Carlo algorithm. Applying our approach to a 3D Yukawa Fermi liquid, we demonstrate that the renormalized perturbation theory remains predictive even in the strongly correlated regime and uncover significant sign cancellations between different channels contributing to the scattering amplitude. Our work establishes a novel framework for investigating strong correlations in quantum many-body systems, offering a systematic approach to explore nonlocal theories for challenging systems like the electron liquid in material science. 
\end{abstract}
\maketitle


\section{Introduction}

Emergent quantum fields in many-body systems, characterized by distinct spatial and temporal structures within the Newtonian framework, inherently feature nonlocal effective interactions. These interactions naturally arise in the long-wavelength and low-energy limit, exhibiting intricate patterns that extend beyond the local, gauge boson-mediated forces described by conventional quantum field theory (QFT)~\cite{dual_fermion, dual_fermion2, fRG_review}. A prime example where nonlocal QFT can be highly relevant is the Fermi liquid, a fundamental quantum many-body state that describes interacting fermions in diverse systems, including the conduction electrons in real materials \cite{Giuliani_Vignale_2005}, ultracold atomic gases \cite{Zhai_2021,cold_atom}, and neutron star matter \cite{neutron_star_matter}. Near the Fermi surface, quasiparticles with renormalized dispersions and interactions govern the low-energy dynamics of Fermi liquids ~\cite{FL1,FL2}. Deriving these renormalized properties from the bare theory poses a significant challenge, as many candidate field theories for Fermi liquids feature nonlocal interactions in space and time rather than purely local ones \cite{polchinski2, shankar, hewson_RG, Dupuis, Lee_FL_1, Lee_FL_2}. Developing a systematic field-theoretical framework to solve these nonlocal QFTs is crucial for tackling the Fermi liquid problem and enabling precise predictions of low-energy properties from first principles.


The primary challenge in nonlocal QFTs, known as the ``vertex problem", stems from the difficulty in systematically resumming the full nonlocal interaction vertex within the framework of many-body perturbation theory. This resummation process, known as the renormalization of interactions, is crucial for obtaining a well-behaved perturbative expansion that accurately captures the low-energy physics. Existing quantum many-body approaches, such as Dyson-Schwinger equations~\cite{dyson, schwinger}, functional renormalization group equations (fRG)\cite{fRG2, fRG3, fRG_review}, and parquet equations\cite{parquet1, parquet2, parquet3}, attempt to tackle this challenge by employing a skeleton diagrammatic framework to resum the full interaction vertex. However, despite their successes in various applications, these methods encounter significant obstacles that lie at the heart of the vertex problem.

The vertex problem manifests itself in several ways. Firstly, the full interaction vertex exhibits a complex dependence on multiple external momenta and frequencies, leading to a high-dimensional parameter space that is computationally challenging to handle, a phenomenon known as the curse of dimensionality. This complexity is particularly pronounced in strongly correlated systems, such as the Hubbard model near half-filling and quantum critical points. In these systems, the interaction vertex develops intricate structures with multiple singularities~\cite{hubbard_vertex1, hubbard_vertex2, hao1975, hao2014skeleton}, arising from the interplay between competing ordering tendencies and the presence of low-energy collective modes. Capturing these rich features necessitates a high-dimensional representation of the vertex, which poses significant computational challenges. Secondly, the full interaction vertex obtained from skeleton diagrammatic techniques often violates fundamental conservation laws or Ward identities~\cite{Baym1, Baym2}, leading to unphysical results that undermine the reliability of the calculations. This issue is particularly troublesome, as it suggests that the resummation of the full vertex may not always yield a consistent and physically meaningful theory. Finally, the skeleton diagrammatic series may fail to converge to the correct answer, especially in strongly correlated systems where the perturbative expansion is not well-controlled~\cite{kozik2015}. This raises concerns about the reliability and accuracy of the results obtained from these methods, even when the full vertex can be successfully resummed. Taken together, these issues constitute the vertex problem and pose a significant barrier to the accurate description of nonlocal interactions in quantum many-body systems. Overcoming the vertex problem is therefore crucial for unlocking the full potential of nonlocal QFTs in describing the rich physics of strongly correlated systems.

Recent numerical efforts to mitigate the vertex problem have focused on finding efficient representations that compress the complex space-time structure of the vertex function into a manageable number of basis functions. This idea has been particularly successful in the imaginary-time domain, where universally optimized basis functions have been rigorously derived from the Lehmann representation of Green's functions. Notable examples include the intermediate representation (IR)\cite{IR1, IR2} and the discrete Lehmann representation (DLR)\cite{DLR}. However, finding efficient representations that capture both the spatial and temporal structure of the vertex function remains a significant challenge. Two promising approaches that have emerged are the partial bosonization of the vertex function and the use of tensor network approximations. In the partial bosonization approach~\cite{hubbard_basis1, hubbard_basis2, hubbard_basis3}, the vertex function is decomposed into single-boson exchange contributions, reducing the complexity of diagrammatic expansions and improving the convergence properties of the resulting equations. However, a key challenge in this approach is finding an accurate parameterization of the irreducible vertex function that cannot be directly bosonized, which is crucial for capturing the full complexity of the vertex function, especially in strongly correlated regimes. Tensor network representations, such as the tensor train approximation~\cite{tensor_train1, tensor_train2}, have also shown promise in tackling the vertex problem by exploiting the low-rank structure of the vertex function in the space-time domain. While tensor networks offer a compact and efficient representation, the effectiveness and accuracy of this approach across a wide range of physical systems and parameter regimes remains an open question. Despite the progress made in developing efficient representations, finding a universal and robust approach that can accurately capture the full space-time structure of the vertex function while maintaining computational feasibility remains an ongoing challenge. The limitations of existing methods highlight the need for new and innovative approaches that can simultaneously achieve accuracy and efficiency in the resummation of the vertex function. 

In this paper, we propose a systematic approach to tackle the vertex problem by selectively resumming only the essential part of the vertex interaction while treating the remaining corrections perturbatively. This strategy is based on the insight that the essential part, which captures the most important low-energy physics, is often much simpler to represent than the full vertex function. By focusing on this essential part, identified through physical considerations or approximated using suitable representations, we can circumvent the computational complexity associated with the full vertex while still accurately describing the key physical processes. Our theory allows for the systematic computation of corrections using renormalized Feynman diagrams, ensuring that even a crude initial approximation can serve as a reliable starting point.


Our scheme goes beyond a mere technical development; it generalizes the well-established field-theoretical renormalization scheme~\cite{peskin2018introduction,bphz2,bphz3,bphz4}, originally developed for local QFTs in high-energy physics~\cite{QED1,QED2,QED3}, to nonlocal QFTs in many-body physics. The cornerstone of our approach is a momentum-resolved renormalization condition based on projection operators, which extends the concept of coupling constant renormalization to nonlocal interactions.  This generalized condition gives rise to renormalized Feynman diagrams and counterterms that are functionals of the nonlocal interactions. To systematically derive these diagrams, we introduce a parquet algorithm that efficiently constructs high-order vertex diagrams by combining lower-order subdiagrams. Moreover, we develop a dedicated diagrammatic Monte Carlo algorithm (DiagMC)~\cite{diagMC1,diagMC2,diagMC3,diagMC4,boldDiagMC, sce1, leblanc} specifically designed to efficiently compute the resulting diagrammatic series. By generalizing the field-theoretical renormalization scheme to nonlocal QFTs, our approach opens up an alternative avenue for addressing the vertex problem.

The practical strength of our approach is showcased in its application to a three-dimensional (3D) Fermi liquid system with a Yukawa interaction. By focusing on the calculation of the 4-vertex function, including the scattering amplitude and the Landau quasiparticle interaction, we derive the precise quasiparticle effective interaction from the microscopic theory. Our results show that the renormalized perturbation theory remains predictive even in the strongly correlated regime where the bare perturbation theory fails.

Remarkably, our work uncovers significant cancellations between different channels contributing to the scattering amplitude, providing a deeper understanding of how Fermi liquids maintain their stability and exhibit emergent weakly interacting behavior despite strong bare couplings. First, as predicted by the conventional wisdom of particle-hole screening, we observe a substantial cancellation between the bare coupling and the particle-hole channel, indicating a strong screening effect. Second, and more remarkably, we find a strong cancellation between the particle-hole-exchange channel and the particle-particle channel, which further reduces the scattering amplitude. Interestingly, we also find that the interaction vertex function exhibits a relatively weak angle dependence, suggesting emergent locality in the quasiparticle interaction, despite the momentum-dependence of the bare interaction. These findings highlight the importance of considering the full complexity of many-body effects when studying strongly correlated systems and provide valuable insights into the mechanisms responsible for the emergent properties of Fermi liquids.

The remainder of this paper is structured as follows. Sec. II introduces a nonlocal quantum field theory for Fermi liquids and discusses the challenges posed by the vertex problem in many-body perturbation theory. We also review the field-theoretical renormalization scheme in local QFT and explore its potential generalization to nonlocal theories. Sec. III presents our projection-based renormalization condition that selectively resums the essential nonlocal contributions to the effective interaction vertex. Sec. IV develops a renormalized Feynman diagrammatic expansion, introducing an algorithm for generating high-order diagrams using the parquet equations. Sec. V  discusses the importance of the imaginary-time representation and proposes a projection operator design that mitigates the sign problem, which enables an efficient DiagMC algorithm for evaluating the renormalized diagrams. Sec. VI demonstrates the application of our approach to a 3D Yukawa Fermi liquid, showcasing its predictive power in the strongly correlated regime and uncovering significant sign cancellations between different scattering channels. Sec. VII summarizes our findings, discusses broader implications, and outlines future directions.

\section{Motivation}
\subsection{Nonlocal Field Theory of Fermi liquids}
We investigate a many-fermion system with generic dispersion and nonlocal two-body interactions, providing a versatile model for various physical phenomena. The system's bare action is described by:
\begin{equation}
\label{eq:model}
S = \int_{12} \psi^\dagger_{1}g^{-1}_{12}\psi_{2} + \frac{\xi}{4}\int_{1234} u_{1234}\psi^\dagger_{1}\psi^\dagger_{2}\psi_{3}\psi_{4},
\end{equation}
Here, $\psi^\dagger$ and $\psi$ are Grassmann variables, with indices $i = 1, 2, 3, 4$ representing space/imaginary-time coordinates and the spins $i = (\vec{x}_i, \tau_i, \sigma_i)$. The bare fermionic propagator is denoted by $g_{12}$, while the bare interaction term $u_{1234}$ is short-ranged but can exhibit complex space-time behavior. We introduce a constant $\xi=1$ to track the perturbation order. This interaction term encompasses a wide range of physical scenarios, such as the Yukawa potential, which plays a fundamental role in describing nuclear forces~\cite{nuclear_force}, and the statically screened Coulomb interaction, a key component in understanding the dynamics of electrons in metals~\cite{Giuliani_Vignale_2005}. By setting the interaction strength to an intermediate level, we ensure that the system remains within the Fermi liquid regime while exhibiting nontrivial renormalization effects. In this regime, the low-energy dynamics of the system is controlled by quasiparticles with renormalized mass and interactions. The versatility of this model makes it applicable to various fields, including condensed matter physics, nuclear physics, and ultracold atomic gases, providing a unified framework for studying many-body phenomena across different scales and systems.

To address the quasiparticle dynamics near the Fermi surface, which may differ significantly from those of bare particles, we introduce an effective action that incorporates renormalization effects. This effective action, given by,
\begin{equation}
\label{eq:renormalized_theory}
    S= \int_{12} \bar{\psi}^\dagger_{1}G^{-1}_{12}\bar{\psi}_{2}+\frac{\xi}{4}\int_{1234} R_{1234}\bar{\psi}^\dagger_{1}\bar{\psi}^\dagger_{2}\bar{\psi}_{3}\bar{\psi}_{4}+\textrm{CTs}.
\end{equation}
The quasiparticle propagator $G$ and the screened interaction $R_{1234}$  are crucial for capturing the renormalization of the bare propagator and interactions caused by the particle-hole fluctuations. The quasiparticle fields are represented by ${\bar{\psi}}^\dagger$ and $\bar{\psi}$. Although the theory allows for more complex multi-quasiparticle interactions, modern RG analyses suggest that these are irrelevant at the Fermi liquid fixed point\cite{shankar, hewson_RG, Dupuis, fRG_review}. Consequently, we assume that these interactions are only perturbatively significant and do not include them at the tree level in our model.

The renormalized parameters in the effective action, such as the quasiparticle propagator and the screened interaction, already incorporate a resummation of many-body effects. To avoid double-counting these effects in the perturbation theory, we introduce a set of counterterms (CTs) in Eq. (\ref{eq:renormalized_theory}). They effectively subtract out the many-body contributions that are already accounted for in the renormalized parameters, ensuring that the perturbative expansion remains consistent and accurate. The specific form of the counterterms should be determined by the definition of the renormalized coupling $R$, which will be discussed in the following sections.

In general, the renormalized coupling $R$ in the action is nonlocal due to the system's characteristic length scales $1/k_F$ and energy scales $E_F$. The tools and techniques developed for local QFT cannot be directly applied to nonlocal interactions, presenting a significant challenge in the study of these systems.

\subsection{Field-theoretical Renormalization Scheme}

To develop a framework for solving nonlocal QFTs, it is instructive to review the field-theoretical renormalization scheme~\cite{peskin2018introduction,bphz2,bphz3,bphz4}, which has been remarkably successful in tackling local QFTs such as quantum electrodynamics (QED)~\cite{QED1,QED2,QED3}. In local QFTs, bare perturbation theory often proves inadequate due to the presence of large parameters arising from the renormalization of interactions. These large parameters typically manifest as ultraviolet (UV) divergences in loop integrals, rendering the perturbative expansion ill-defined. The field-theoretical renormalization scheme systematically addresses this issue by introducing counterterms in the action. These counterterms are determined by imposing renormalization conditions that establish a well-defined relationship between the bare and renormalized quantities at a specific energy scale, known as the renormalization scale. Although condensed matter systems generally do not suffer from UV divergences due to the presence of natural cutoffs such as the lattice spacing, the challenge of dealing with large parameters arising from the renormalization of interactions remains pertinent.


A key feature of the field-theoretical renormalization scheme is that it focuses on renormalizing only the most relevant coupling constant in the interaction vertex, while treating the remaining part, which may exhibit a complex dependence on external momenta and frequencies, perturbatively using renormalized Feynman diagrams. This approach circumvents the need for an explicit calculation of the full effective interaction with all its intricate details, as only the renormalization of the most relevant part needs to be tracked. Although this scheme has proven highly effective in the context of local QFTs, its generalization to nonlocal interactions has not yet been thoroughly explored in the realm of nonlocal QFTs in many-body problems.

In this work, we delve into a renormalization scheme that deals with nonlocal interactions. Our approach involves two steps: 1) We will develop an ansatz for the renormalized coupling $R$ that accurately captures its momentum and frequency dependence. To accomplish this, we need to go beyond the standard local QFT framework and take into account the specific features of quantum many-body systems, such as the Fermi surface and its associated length and energy scales. 2) We will adapt the standard renormalized diagrammatic technique of local QFT to nonlocal theories. This will require us to develop efficient algorithms to derive and compute the diagrams, as well as the counterterms. By tackling these challenges, we seek to extend the powerful tools of renormalized field theory to a wider range of quantum many-body systems, deepening our understanding of these complex systems and enabling more accurate predictions of their low-energy properties.

\begin{figure}[!htbp]
\centering
\includegraphics[scale=0.4,angle=0,width=0.9\columnwidth]{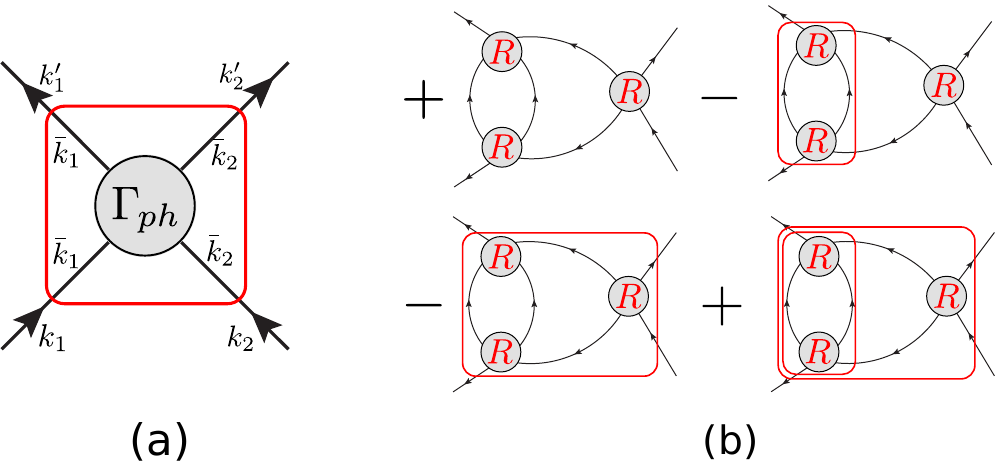}
\caption{\label{fig:R_projection} a) The projection operator (the red box) to extract the forward-scattering part of the particle-hole-reducible vertex function. The incoming momenta/frequencies $k_1$ and $k_2$ are projected onto the Fermi surface ($k=(\vk, i\omega_n) \rightarrow \bar{k}=(k_F\vk/|\vk|, i\pi T)$), and two outgoing momenta/frequencies are projected to the EXACT forward-scattering direction. b) Examples of the third-order renormalized diagrams of the vertex function $\Gamma_4$. It demonstrates the cancellation between the counter-diagrams (in the red box) and the corresponding vertex subdiagrams. }
\end{figure}

\section{Projection-based Renormalization Condition}
To address the first challenge of developing an ansatz for the renormalized coupling $R$, we propose a projection-based renormalization condition. In QFT, renormalization conditions define the relationship between bare and renormalized quantities, such as couplings and fields, at a particular energy scale~\cite{peskin2018introduction}. These conditions are essential for ensuring the perturbation theory yields physically meaningful results. However, the traditional local QFT approach of defining renormalized couplings as constants at a specific energy scale is insufficient for Fermi liquids due to the nonlocal nature of their interactions.

We focus on the 4-point vertex function, $\Gamma_4(k_1, k_2; k'_1, k'_2)$, which depends on four momentum-frequency vectors: two incoming particles with $k_1$ and $k_2$, and two outgoing particles with $k'_1$ and $k'_2$. Due to the momentum-frequency conservation law, $k_1 + k_2 = k'_1 + k'_2$, only three of these four vectors are independent. The 4-point vertex function $\Gamma_4$ consists of two parts: the bare interaction $u_{k_1k_2; k'_1k'_2}$ and the quantum many-body correction $\Delta \Gamma_4(k_1, k_2; k'_1, k'_2)$. The bare interaction $u$ possesses a simple analytic form and captures the essential UV physics. In contrast, $\Delta \Gamma_4$ exhibits a complex dependence on momentum and frequency encodes the emergent infrared (IR) physics. 

In the context of Fermi liquids, the quantum many-body correction $\Delta \Gamma_4$ receives significant contributions from scattering processes mediated by particle-hole excitations, which can be further decomposed into the direct particle-hole channel $\Gamma_{ph}$ and the exchanged particle-hole channel $\tilde{\Gamma}_{ph}$. Diagrammatically, $\Gamma_{ph}$ represents particle-hole reducible diagrams, which are Feynman diagrams that can be divided into two separate diagrams by cutting one particle and one hole propagator line. On the other hand, $\tilde{\Gamma}_{ph}$ diagrams are derived from $\Gamma_{ph}$ diagrams by exchanging the two outgoing external legs. These two channels exhibit distinct low-energy behaviors. The direct particle-hole channel $\Gamma_{ph}$ is characterized by non-analytic scattering amplitudes when the momentum-frequency transfer $q = k_1 - k'_1$ is small, signaling a soft particle-hole excitation. In contrast, the exchanged particle-hole channel $\tilde{\Gamma}_{ph}$ features non-analyticity when the exchanged momentum-frequency transfer $\tilde{q} = k_1 - k'_2$ approaches zero. Elucidating the contributions from these channels is crucial for capturing the essential physics of particle-hole excitations and their role in renormalizing the effective interactions in Fermi liquids.

To incorporate the most relevant physics from both components, we define the renormalized interaction $R$ as:
\begin{equation}
\label{renorm_cond}
R \equiv u + \mathcal{P} \Delta \Gamma_4.
\end{equation}
where $\mathcal{P}$ represents a projection operator applied to $\Delta \Gamma_4$, 
\begin{eqnarray}
\label{eq:projection}
    \mathcal{P}\Delta \Gamma_4 (k_1, k_2; k_1'; k_2')&= \Gamma_{ph}(\bar{k}_1, \bar{k}_2; \bar{k}_1, \bar{k}_2)\cdot e^{-\bold{q}^2/2\delta}+ \nonumber \\
    &\tilde{\Gamma}_{ph}(\bar{k}_1, \bar{k}_2; \bar{k}_2, \bar{k}_1)\cdot e^{-\bold{\tilde{q}}^2/2\delta}.
\end{eqnarray}
 
 The projection operator $\mathcal{P}$ maps the incoming and outgoing momenta/frequencies ($k_1, k_2$ and $k_1', k_2'$) onto the Fermi surface, with $\bar{k}_i$ denoting the projected momenta/frequencies (Fig. \ref{fig:projection}(a)). This simplifies the quantum correction to primarily depend on the angles between the incoming and outgoing momenta. The small parameter $\delta$ satisfies $\delta \ll k_F$, where $k_F$ is the Fermi momentum, ensuring that the projection operator focuses on the forward-scattering contributions, which are crucial for the low-energy behavior of Fermi liquids~\cite{shankar}. 
 
By constructing $R$ as a sum of the bare interaction $u$ and the projected quantum correction $\mathcal{P} \Delta \Gamma_4$, our renormalization condition (Eq. (\ref{renorm_cond})) interpolates the bare interaction in the UV limit to the quasiparticle interaction in the IR limit, capturing the most relevant physics at different scales. Following the seminal work by Shankar~\cite{shankar}, we focus on the momentum dependence and omit the frequency dependence in our renormalization condition, effectively treating $R$ as a static interaction. 


\section{Renormalized Feynman Diagrammatic Expansion}
The projection-based renormalization condition in Eq. (\ref{renorm_cond}) lays the foundation for a renormalized perturbation theory of the 4-vertex function $\Gamma_4$, which is central to describing quasiparticle scattering amplitudes. 

In Fermi liquids with strong bare interactions, straightforward perturbation theory based on the bare coupling $u$ fails to capture the essential physics due to significant higher-order contributions arising from the renormalization of both the single-particle propagator and the two-particle interaction vertex (e.g., 4-vertex function). The renormalization of the propagator can be addressed using the well-established skeleton diagrammatic approach, where the bare propagator is replaced by the renormalized quasiparticle propagator $G$, and all self-energy insertions are eliminated. For brevity, we will assume that this renormalization has been performed and omit the explicit dependence on $G$ in the subsequent discussion. However, the renormalization of the interaction vertex poses a more formidable challenge, as discussed in the introduction, leading to the notorious vertex problem. To tackle this issue, we introduce a renormalized perturbation theory based on the renormalized coupling $R$, which effectively captures the essential physics of interaction renormalization while circumventing the computational complexity associated with the full 4-point vertex function.

We start with the bare perturbation theory for $\Gamma_4[u]$, which can be derived using the standard Feynman rules. It's an expansion in powers of the bare coupling $u$:
\begin{equation}
\Gamma_4[u] = u \cdot \xi + \Delta \Gamma_4[u],
\end{equation}
where $\Delta \Gamma_4[u \cdot \xi] = \gamma_2 \cdot \xi^2 + \gamma_3 \cdot \xi^3 + ...$, with $\xi=1$ tracking the number of $u$ in the skeleton Feynman diagrams, and $\gamma_n[u]$ representing the set of diagrams with $n$ instances of $u$. Since $\Delta\Gamma_4$ represents the quantum corrections, its power series in $u$ starts from the second order.

To obtain the renormalized expansion, we re-express the bare coupling $u$ in terms of the renormalized coupling $R$ using the renormalization condition Eq. (\ref{renorm_cond}):
\begin{equation}
u[R] = R \cdot \xi - \mathcal{P}\Delta\Gamma_4[u[R]].
\end{equation}

Substituting this expression into the bare expansion, we derive the renormalized Feynman diagrams for $\Gamma_4$:
\begin{equation}
\Gamma_4[R] \equiv R \cdot \xi + (1 - \mathcal{P})\Delta\Gamma_4[u[R]],
\end{equation}
where $1-\mathcal{P}$ is the projection operator that removes the forward-scattering contributions in the vertex function diagrams. A systematic derivation results in a diagrammatic series as shown in Fig. \ref{fig:R_projection}(b). Remarkably, for every particle-hole loop integral, the renormalized expansion introduces counter-terms that cancel the large contributions from particle-hole fluctuation near the Fermi surface, ensuring a well-behaved perturbative expansion even with strong bare interactions.

\begin{figure}[!htbp]
\centering
\includegraphics[scale=0.4,angle=0,width=0.95\columnwidth]{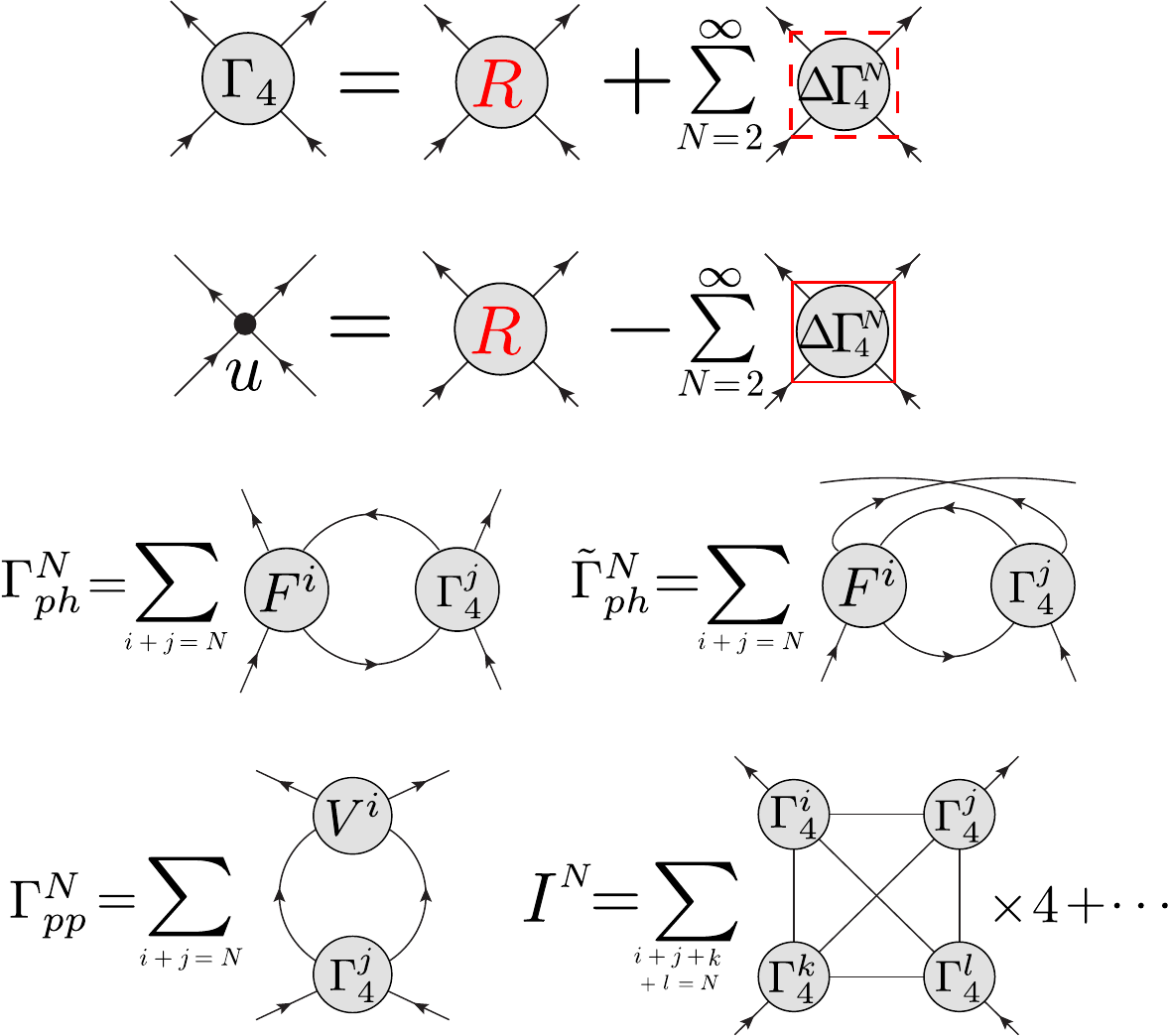}
\caption{\label{fig:iteration} Self-consistent equations for generating high-order renormalized diagrams. The first two equations represent the renormalized expansions for the physical vertex function $\Gamma_4$ and the bare coupling $u$. The red solid box denotes the projection operator $\mathcal{P}$, which extracts the dominant low-energy contribution, while the red dashed box represents $\bar{\mathcal{P}}=1-\mathcal{P}$, which captures the remaining correction. The superscript on each term indicates its order in the perturbative expansion. In this figure, $\Delta \Gamma_4=I+\Gamma_{ph}+\tilde{\Gamma}_{ph}+\Gamma_{pp}$ represents the quantum correction to the physical vertex function, where $I$, $\Gamma_{ph}$, $\tilde{\Gamma}_{ph}$, and $\Gamma_{pp}$ denote the fully irreducible, particle-hole reducible, exchanged particle-hole reducible, and particle-particle reducible contributions, respectively. $F=u+I+\tilde{\Gamma}_{ph}+\Gamma_{pp}$ is the particle-hole irreducible vertex function, and $V=u+I+\Gamma_{ph}+\tilde{\Gamma}_{ph}$ is the particle-particle irreducible vertex function. The fully irreducible vertex $I$ does not contribute until the fourth order.}
\end{figure}

Generating high-order diagrams in the renormalized expansion is a significant challenge. To address this problem, we build upon the algorithm proposed recently in Ref.~\cite{hou2024feynman}, which systematically constructs a Feynman diagrammatic expansion using the parquet equations. These equations, in their conventional form, are a set of self-consistent relations that connect the four-point vertex function to its irreducible components in different channels. The algorithm repurposes the parquet equations by expanding them perturbatively, transforming them into a generator that systematically builds high-order four-point vertex diagrams from lower-order subdiagrams. The resulting high-order Feynman diagrams are not represented as individual diagrams; instead, they are organized into a single comprehensive computational graph of nodes and leaves, where the leaves represent the propagators and interactions, and the nodes represent mathematical operations. This computational graph representation allows for the reuse of many subnodes, greatly simplifying the integrand evaluation process compared to the individual diagram representation. Ref. \cite{Fabian1} has utilized a similar idea to count the number of Feynman diagrams of vertex function.

The original algorithm in Ref.\cite{hou2024feynman}, while effective, does not support the sophisticated nonlocal interaction renormalization scheme required for this work. In that approach, the renormalization of propagators and interactions is implemented through high-order automatic differentiation (AD) of the original Feynman diagrams. This results in renormalized diagrams with counterterms that must be pre-calculated numbers, limiting the scheme's applicability to the more complex nonlocal interaction renormalization employed in this paper. As shown in Fig.\ref{fig:R_projection}, the counterterms in our approach are actually counter-diagrams with internal loop integrations, which is beyond the scope of the method in Ref.~\cite{hou2024feynman}.

To overcome this limitation, we extend the algorithm by introducing a set of renormalized parquet equations, as depicted in Fig.~\ref{fig:iteration}. The equations are derived by re-expanding the bare couplings in the irreducible vertex functions as a power series of renormalized couplings, removing the UV physics in the original bare parquet equations. We then perturbatively expand the renormalized parquet equations to generate the renormalized diagrammatic expansion. The hallmark of our iterative scheme is the simultaneous generation of all diagrammatic topologies and counter-diagrams (namely, the subdiagrams in the red boxes in Fig.\ref{fig:R_projection} and Fig.~\ref{fig:iteration}). Specifically, one starts with the first-order physical vertex function $\Gamma_4^1=R$ and the bare coupling parameterized with the renormalized interaction $u^1=R$, and then substitutes them into the renormalized parquet equations to generate the second-order diagrams of the particle-hole channel $\Gamma_{ph}$, the particle-hole-exchanged channel $\tilde{\Gamma}_{ph}$, and the particle-particle channel $\Gamma_{pp}$. This process is repeated until all diagrams up to order $N$ are generated. Notably, this all-in-one scheme is more straightforward than conventional iterative algorithms such as Bogoliubov's R operation~\cite{bphz1,bphz2,bphz3,bphz4}, where one first needs to list all topologies of bare diagrams and then iteratively generates the corresponding counter-diagrams for each topology.

In contrast to the previous scheme in Ref.~\cite{hou2024feynman}, the counterterms in our approach are not pre-calculated values. Instead, they are represented as projected vertex sub-diagrams that share the same internal variables as their corresponding subdiagrams. This formulation enables the application of more complex renormalization conditions. Furthermore, when the loop integral contains divergences, our approach ensures that the counterterms cancel these divergences before the integral is evaluated. By canceling divergences at the integrand level, we avoid potential numerical instabilities that could arise from directly computing divergent integrals.

 The current algorithm implementation generates fully irreducible diagrams in $I$ by exhaustively considering all possible topology at a given order~\cite{kchen}. For future work, it would be valuable to explore an alternative approach that involves taking the functional derivative of the parquet self-energy diagrams~\cite{Fabian2}.

The renormalized perturbation theory presented here provides a framework for investigating nonlocal interactions in Fermi liquids, enabling accurate calculations of the interaction vertex and other observables. By systematically generating high-order diagrams, we can capture the essential physics of the system while maintaining a well-behaved perturbative expansion, even in the presence of strong bare couplings. 





\section{Imaginary-time Representation}
\label{supply2}

Evaluating the renormalized Feynman diagrams in nonlocal QFT poses significant computational challenges compared to local theories. In local QFT, the effective couplings are typically constants, which allows for the use of analytical techniques like Feynman parameterization to simplify the computation of diagrams. However, in nonlocal QFT, the diagrams are functionals of the renormalized coupling $R(k_1, k_2; k'_1, k'_2)$, which exhibits a complex dependence on the momentum and frequency variables. This functional dependence renders analytical techniques ineffective, necessitating the use of Monte Carlo integration algorithms for diagram evaluation, even at the one-loop level.

The performance of the Monte Carlo integration greatly depends on the sign structure of the diagrammatic series, as sign cancellations between the configurations can result in a severe sign problem, significantly hindering the efficiency and accuracy of the calculation. In our renormalized diagrammatic series, there are two straightforward measures to alleviate the sign problem. First, we should work in the imaginary-time representation instead of the Matsubara-frequency representation, as the propagators and interactions are real-valued in the former, mitigating the sign problem and improving the efficiency of Monte Carlo integration.

Second, to further alleviate the sign problem, we should ensure a massive cancellation between the vertex sub-diagrams and their counter-diagrams at each Monte Carlo step. This raises a crucial question: how can we design the renormalization scheme in Eq.\ref{eq:projection} in the imaginary-time representation to maximize the sign cancellation at the integrand level for any given set of internal and external variables? Although our renormalization scheme fixes the projection operator in the Matsubara-frequency domain, the freedom in its implementation in the imaginary-time domain provides an opportunity to design the projection operator to mitigate the sign problem, making this question essential for the efficiency of our Monte Carlo algorithm.

\begin{figure}[!htbp]
\centering
\includegraphics[scale=0.4,angle=0,width=0.8\columnwidth]{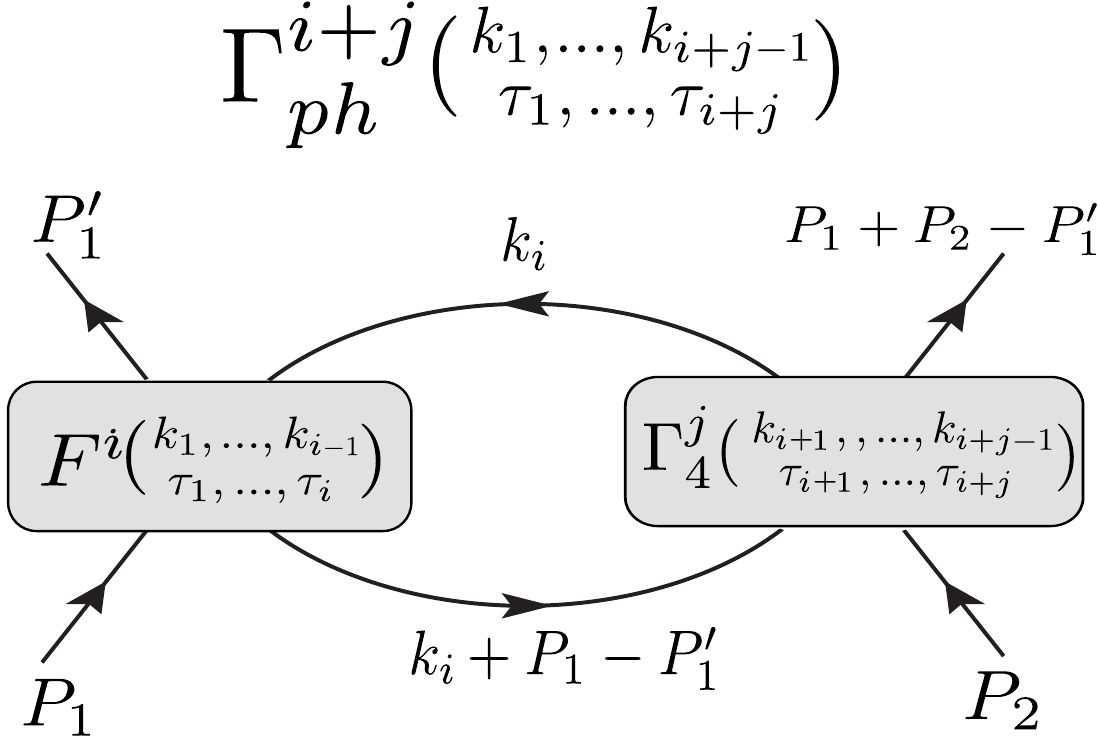}
\caption{\label{fig:iteraction} Recursive relation for constructing an $(i+j)$-order $\Gamma_{ph}$ vertex function from the $i$-order particle-hole irreducible vertex $F$ and the $j$-order full vertex function $\Gamma_4$. In the figure, $k_1, \ldots, k_{i+j-1}$ denote the internal momentum variables, while $P_1$, $P_2$, and $P'_1$ represent the external momentum variables. The temporal variables are labeled as $\tau_1, \ldots, \tau_{i+j+2}$, which include both internal and external variables. The temporal variable of the left incoming leg is fixed to $\tau_1$, while the temporal variables of the other three external legs are determined by the subvertices. This recursive relation allows for the efficient generation of high-order diagrams by combining lower-order subdiagrams. Similar recursive relations can be constructed for the other channels, such as the exchanged particle-hole ($\tilde{\Gamma}_{ph}$) and particle-particle ($\Gamma_{pp}$) channels, enabling a systematic approach to generate all required diagrams in the parquet formalism.}
\end{figure}

\begin{figure}[htbp!]
\centering
\includegraphics[scale=0.4,angle=0,width=0.8\columnwidth]{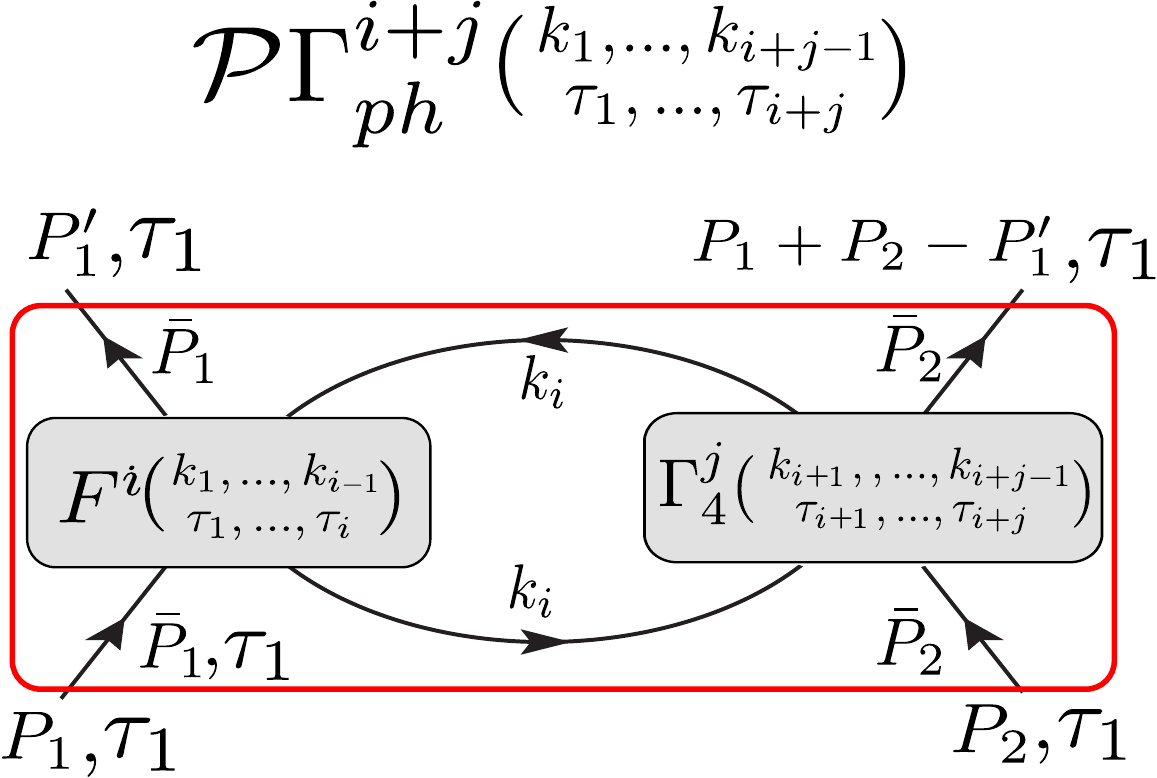}
\caption{\label{fig:projection} Application of the projection operator (red box) to a particle-hole vertex diagram. The operator projects the two incoming momenta onto the Fermi surface, mapping $k \rightarrow \bar{k}$, as explained in the main text. It also projects the temporal variables of all four external legs to the same value as the left incoming leg, ensuring that all external frequencies are projected to $\omega_0 = \pi T$ in the Matsubara frequency representation. This projection operation is essential for isolating the dominant low-energy contributions to the vertex function. It is important to note that the external momenta experience an abrupt change before and after the projection (across the red box). Notably, the projected vertex function still obeys the momentum conservation law. The application of the projection operator is a crucial step in the renormalization procedure, as it allows for the identification and selective resummation of the most important contributions to the vertex function, while avoiding the computational complexity associated with the full vertex function.}
\end{figure}

We propose a projection operator design, as shown in Fig.\ref{fig:projection}, which operators on the vertex sub-diagram in Fig.\ref{fig:iteraction} with the internal variables preserved and aligns all external imaginary-time variables to the same time variable. The operator modifies the external imaginary-time variables to project out the zero-frequency contributions of the vertex sub-diagram, ensuring that the counter-diagram largely balances the vertex sub-diagrams for any given set of internal variables. This design guarantees that the cancellation occurs at the integrand level, before the integration of internal variables takes place, effectively mitigating the sign problem.

Based on this insight, we design a DiagMC algorithm to calculate the diagrams. DiagMC algorithms have emerged as a powerful tool for tackling the computational challenges associated with evaluating Feynman diagrams in various fields of physics~\cite{diagMC1,diagMC2,diagMC3,diagMC4,boldDiagMC, sce1, leblanc}. These algorithms combine the principles of Monte Carlo sampling with the diagrammatic expansion of perturbation theory to efficiently compute high-order contributions to physical observables. There are three main approaches to renormalization in DiagMC: Bold Diagrammatic Monte Carlo (BDMC) replaces bare propagators with fully dressed ones~\cite{boldDiagMC, diagMC3,sce1}, the Renormalized Connected Determinant (CDet) method~\cite{diagMC4, rossi2020renormalized} and combinatorial approach~\cite{kozik2023combinatorial} rely on a determinantal formulation of connected diagrams, and the Taylor-mode automatic differentiation approach represents renormalized diagrams as computational graphs with derivatives computed using generalized chain rules~\cite{kchen, hou2024feynman}.

DiagMC algorithms with renormalized expansions have found numerous applications in quantum many-body physics, yielding state-of-the-art results for various systems. For instance, BDMC has been successfully applied to the Fermi polaron problem~\cite{boldDiagMC, many_polaron}, the Hubbard model~\cite{diagMC3, deng2015emergent, sce2}, the resonant Fermi gas~\cite{van2012feynman, rossi_Unitary}, and the frustrated spin systems~\cite{frustrated_spin_1, frustrated_spin, spin3, spin4}, providing accurate predictions for the equation of state and correlation functions. The CDet method has been employed to calculate the self-energy and spectral functions in the Hubbard model~\cite{diagMC4, cdet_3D_hubbard_half_filling, PhysRevResearch.4.043201, simkovic2022origin} and a variety of other quantum many-body problems~\cite{PhysRevB.101.045134}, while the combinatorial approach has been used to study the equation of state of the 2D SU(N) Hubbard model in an experimentally relevant regime~\cite{kozik2023combinatorial}. Additionally, the Taylor-mode automatic differentiation approach has been utilized to investigate the electron gas problem with long-range Coulomb interactions, yielding precise values for the quasiparticle effective mass and other low-energy quantities~\cite{kchen,haule2022single, wdm, UEG_dynamic, hou2024feynman}.

Despite these successes, existing DiagMC algorithms have primarily focused on the renormalization of single-particle properties (of fermions or force-mediating bosons), leaving the computation of renormalized Feynman diagrams with generic non-local interaction counter-diagrams as an unexplored frontier. The challenges associated with the functional dependence of the renormalized coupling on momentum and frequency variables, as well as the presence of counter-diagrams, have hindered the development of efficient DiagMC algorithms for this problem.

To address this challenge, we have developed an efficient diagrammatic Monte Carlo (DiagMC) algorithm that tackles the high-order integrals by using the renormalized diagrams of nonlocal QFT. The primary objective of this algorithm is to calculate the order-$N$ contribution of the renormalized vertex function $\Gamma^{(N)}_4$ for a given set of external variables. Mathematically, the $N$th-order contribution we aim to evaluate is a high-dimensional integral,
\begin{equation}
\Gamma^{(N)}_4=\int dk^{N-1} d\tau^N \sum_{t \in \mathrm{T}} W_t(k_1, \ldots, k_{N-1}; \tau_1, \ldots, \tau_{N}),
\end{equation}
where $W_t$ is an $N$th-order renormalized diagram with the topology label $t$, involving $N-1$ internal momenta $(k_1, k_2, \ldots, k_{N-1})$ and $N$ internal imaginary-time variables $(\tau_1, \tau_2, \ldots, \tau_N)$. 

Conventional DiagMC methods that sample the diagram topology face significant challenges when adapted to our renormalized Feynman diagrams. In these approaches, diagram topology and internal variables are sampled stochastically, performing a random walk with a distribution $\propto \sum_{t \in \mathrm{T}} |W_t|$. However, when two diagrams differ by the replacement of a vertex sub-diagram with a counter-diagram, they experience substantial sign cancellation. Sampling the absolute sum of these two diagrams leads to a severe sign problem, limiting the computational accuracy and efficiency of conventional DiagMC. To overcome this problem, we have developed a DiagMC algorithm specifically designed for renormalized Feynman diagrams with nonlocal interactions. Instead of sampling the diagram topology, our approach samples the total summed weights $\propto |\sum_{t \in \mathrm{T}} W_t|$ of all diagrams and counter-diagrams at the same order, which can be efficiently evaluated using the parquet formalism. This enables the cancellation of vertex sub-diagrams and their counter-diagrams in each Monte Carlo step, effectively mitigating the sign problem.

By leveraging the advantages of the imaginary-time representation, carefully designing the projection operator, and developing a DiagMC algorithm, we can efficiently compute high-order contributions to the renormalized vertex function, paving the way for a deeper understanding of the rich physics in nonlocal QFTs.

\section{Application}
To demonstrate the effectiveness of our renormalization scheme, we apply it to a 3D Fermi liquid with a Yukawa interaction.  To focus on the effect of interaction renormalization, we consider a simplified model where the self-energy renormalization is treated analogously to the renormalized perturbation theory in QED. In QED, the details of the bare electron propagator (at the Planck scale) are not important; only the renormalized electron propagator measured at low energies appears in the renormalized diagrams. Similarly, in our model, the bare fermionic propagator is fine-tuned such that the fully dressed Green's function takes the form $G_k=-1/(i\omega_n-{\bold{k}}^2/2m+E_F)$, removing all self-energy insertions from the diagrams. The system is characterized by the Wigner-Seitz radius $r_s$, which determines the Fermi energy $E_F=(9\pi/2S)^{2/3}/r_s^2$ in atomic Rydberg units, where $S=1$ for the polarized case and $S=2$ for the unpolarized case. The bare interaction is the statically screened Coulomb repulsion $u_{qkk'}=8\pi /({\bold{q}}^2+\lambda)$, with $\lambda$ as the inverse screening length. Calculations are performed at $T=0.025E_F$.

The renormalized coupling is derived from the bare coupling by solving a self-consistent equation:
\begin{equation}
R = u+\mathcal{P}\Delta F[u[R]] + \mathcal{P}\left[(u+\Delta F[u[R]])GG \Gamma_4[u[R]]\right].
\end{equation}
Here, $\Delta F\equiv \Delta \Gamma_4-\Gamma_{ph}$ represents the particle-hole irreducible (PHI) diagrams. It includes all the 4-vertex function diagrams except for the bare interaction $u$ and the particle-hole reducible diagrams. By excluding the particle-hole reducible diagrams, $\Delta F$ captures the essential many-body correlations without double-counting the contributions from particle-hole channels. Both $\Delta F[u[R]]$ and $\Gamma_4[u[R]]$ are perturbatively expanded in powers of $R$. We solve this self-consistent equation with a DiagMC to derive the renormalized parameters from the bare coupling. It is also possible to derive the renormalized parameters using the functional RG\cite{wilson1, fRG_review} or the field-theoretic RG\cite{RG_origin}. For simplicity, we will not discuss the detail here.
\begin{figure}[htbp]
\centering
\includegraphics[scale=0.4,angle=0,width=1.0\columnwidth]{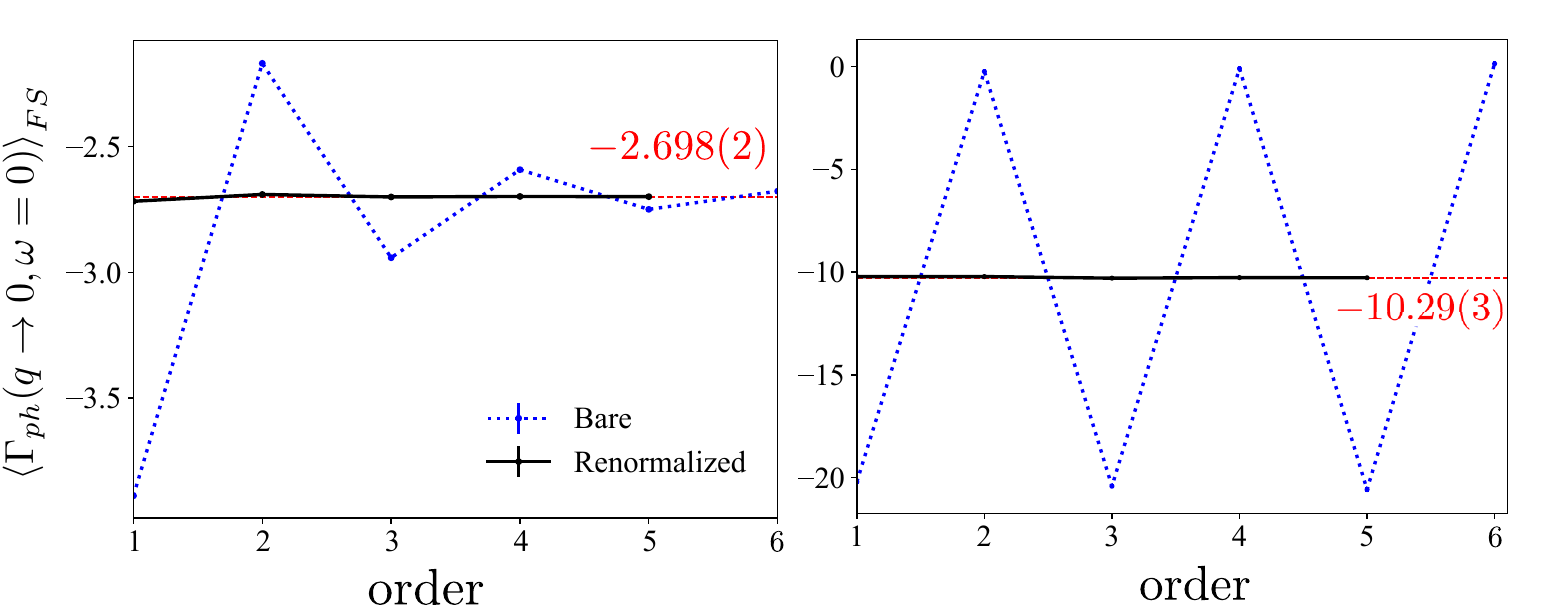}
\vspace{-5mm}
\caption{\label{fig:convergence}  Comparison of the forward-scattering particle-hole vertex function, averaged on the Fermi surface, calculated using the bare expansion (blue) and the renormalized expansion (black). The loop order of the expansion is denoted by $n$. The calculations are performed for two polarized Yukawa Fermi liquids with a Wigner-Seitz radius of $r_s=2^{1/3}$ and inverse screening lengths of $\lambda=2$ (left panel) and $\lambda=1$ (right panel). In the system with stronger screening ($\lambda=1$), the bare expansion diverges, indicating the breakdown of the perturbative approach. In contrast, the renormalized expansion remains well-behaved and predictive, demonstrating the effectiveness of the renormalization scheme in capturing the essential physics of the strongly correlated system. }
\end{figure}

The renormalization's effectiveness is demonstrated by calculating the static particle-hole vertex function with renormalized and bare expansions in the fully polarized limit (Fig. \ref{fig:convergence}). The renormalized expansion is significantly more predictive than the bare expansion.

To investigate the impact of strong correlations, we study the two-body scattering amplitude and the Landau quasiparticle interaction of an unpolarized Yukawa Fermi liquid with $r_s=4$ and $\lambda=k_F^2$. The scattering amplitude $\Gamma^q$ is defined as the 4-vertex function averaged on the Fermi surface in the static limit with a vanishingly small momentum transfer,
\begin{equation}
\Gamma^q(\theta) \equiv R(q \rightarrow 0, \theta)\equiv \frac{1}{N_F} \left[A^s(\theta) \cdot I+A^a(\theta) \vec{\sigma} \cdot \vec{\sigma}\right],
\end{equation}
where $\theta$ is the angle between the two incoming momenta, and $N_F$ is the density of states on the Fermi surface. In the paramagnetic limit, the scattering amplitude is decomposed into a spin-symmetric component $A^s$ and a spin-asymmetric component $A^a$. Each component can be further decomposed into different angular momentum channels using the Legendre transform $A^{s,a}(\theta) = \sum_{l=0}^{\infty} (2l+1) A^{s,a}_l P_l(\cos \theta)$, where $A^{s,a}_l$ are the Legendre coefficients and $P_l(\cos \theta)$ are the Legendre polynomials. The Landau quasiparticle interaction $F^{s,a}(\theta)$, on the other hand, is defined as the 4-vertex function in the zero transfer momentum and small transfer frequency limit, averaged over the Fermi surface. It is related to the scattering amplitude by the relation: $A_l^{s, a}=F_l^{s, a} /\left(1+F_l^{s, a}\right)$, where $F_l^{s,a}$ are the Legendre coefficients of $F^{s,a}(\theta)$.  

\begin{figure}[htbp]
\centering
\includegraphics[scale=0.4,angle=0,width=0.8\columnwidth]{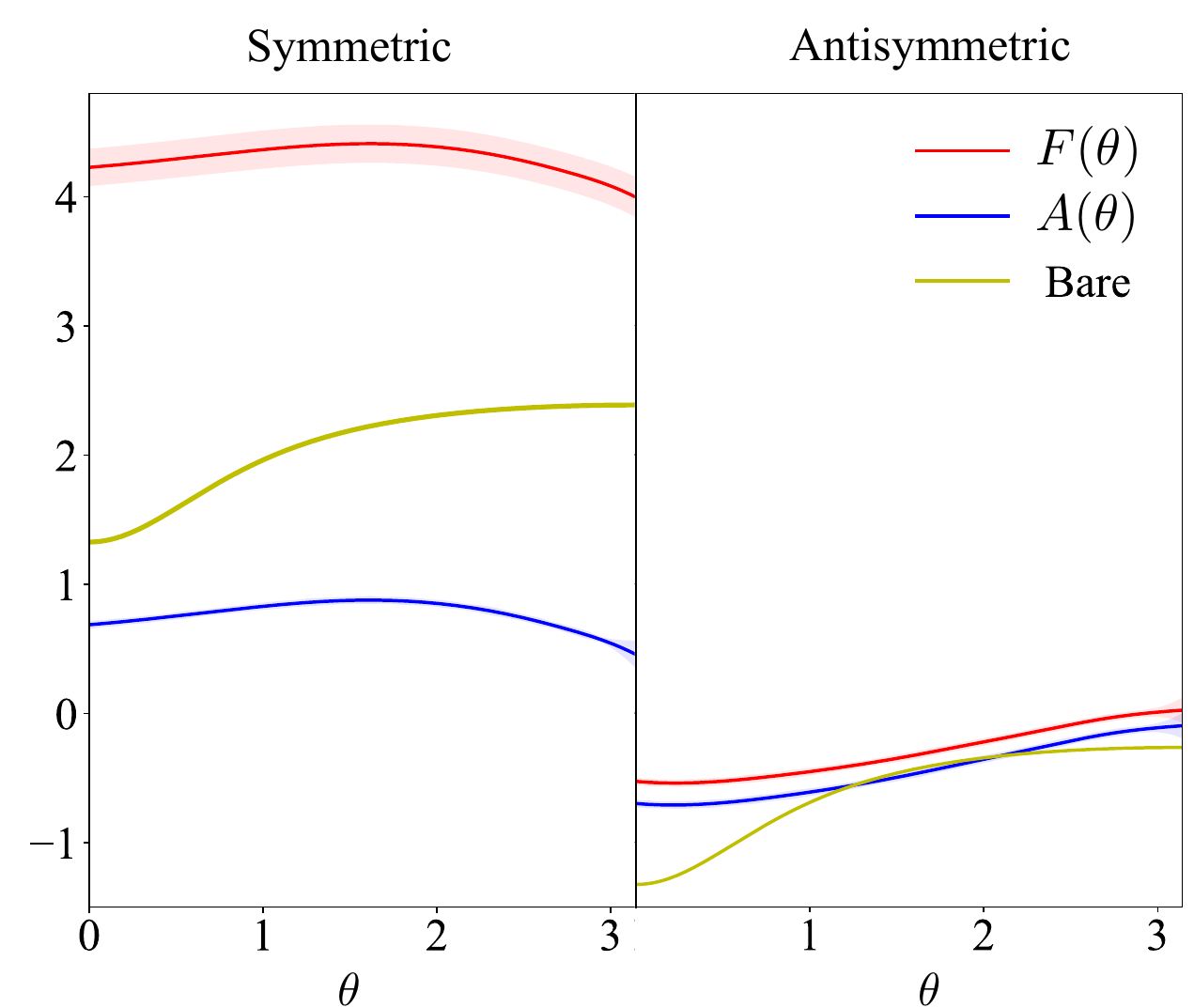}
\caption{\label{fig:vertex} Scattering amplitude (blue) and the quasiparticle interaction (red) as a function of the angle $\theta$ between two incoming momenta on the Fermi surface, for an unpolarized Yukawa Fermi liquid with $r_s=4$ and $\lambda=k_F^2$. The scattering amplitude is evaluated at zero transfer momentum and frequency, whereas the quasiparticle interaction amplitude is assessed at zero transfer frequency. The left and right panels show the spin-symmetric and spin-antisymmetric components, respectively, normalized against the Fermi surface's density of states. Compared to the bare interaction (yellow solid curves), both the quasiparticle interactions and the scattering amplitudes exhibit weaker angular dependence, indicating emergent locality.}
\end{figure}

Accurate results for the scattering amplitude and the Landau quasiparticle interaction as a function of the angle between the incoming momenta are presented in Fig. \ref{fig:vertex}. These quantities, which represent different momentum-frequency slices of the full 4-vertex function, exhibit weak angle dependence, suggesting that the 4-vertex function is relatively local. This locality is a consequence of the many-body corrections and the screening effects, which significantly modify the momentum and frequency dependence of the bare interaction. 

\begin{table}[h]
\centering
\begin{tabular}{c| c c|c c }
\hline
$l$ & $A^s_l$ & $F^s_l$ &$A^a_l$ & $F^a_l$ \\
\hline
0 & 0.81(1)  & 4.3(3) & -0.460(8) & -0.315(4)\\
1 & 0.003(1) &  0.003(1) & -0.032(2) & -0.031(2)\\
2 & -0.0057(8) & -0.0055(8) & 0.0014(8) & 0.0014(8)\\
\hline
\end{tabular}
\caption{Measured Legendre coefficients of the scattering amplitude ($A_l$) and the Landau quasiparticle interaction ($F_l$) for an unpolarized Yukawa Fermi liquid with $r_s=4$ and $\lambda=k_F^2$. }
\label{tab:legendre_coefficients}
\end{table}

To quantitatively analyze the amplitude of the scattering amplitude and the Landau quasiparticle interaction, we present their Legendre coefficients, $A_l^{s,a}$ and $F_l^{s,a}$, in Table \ref{tab:legendre_coefficients}.  The coefficients for $l>1$ are vanishingly small, providing further evidence for the emergent locality of the four-vertex function. A striking observation is that the spin-symmetric Landau quasiparticle interaction is significantly larger than the bare coupling, with $F_0^s$ reaching a value of $4.3(3)$. This substantial enhancement highlights the strong many-body renormalization of the quasiparticle interaction in the system.

\begin{figure}[htbp]
\centering
\includegraphics[scale=0.4,angle=0,width=0.8\columnwidth]{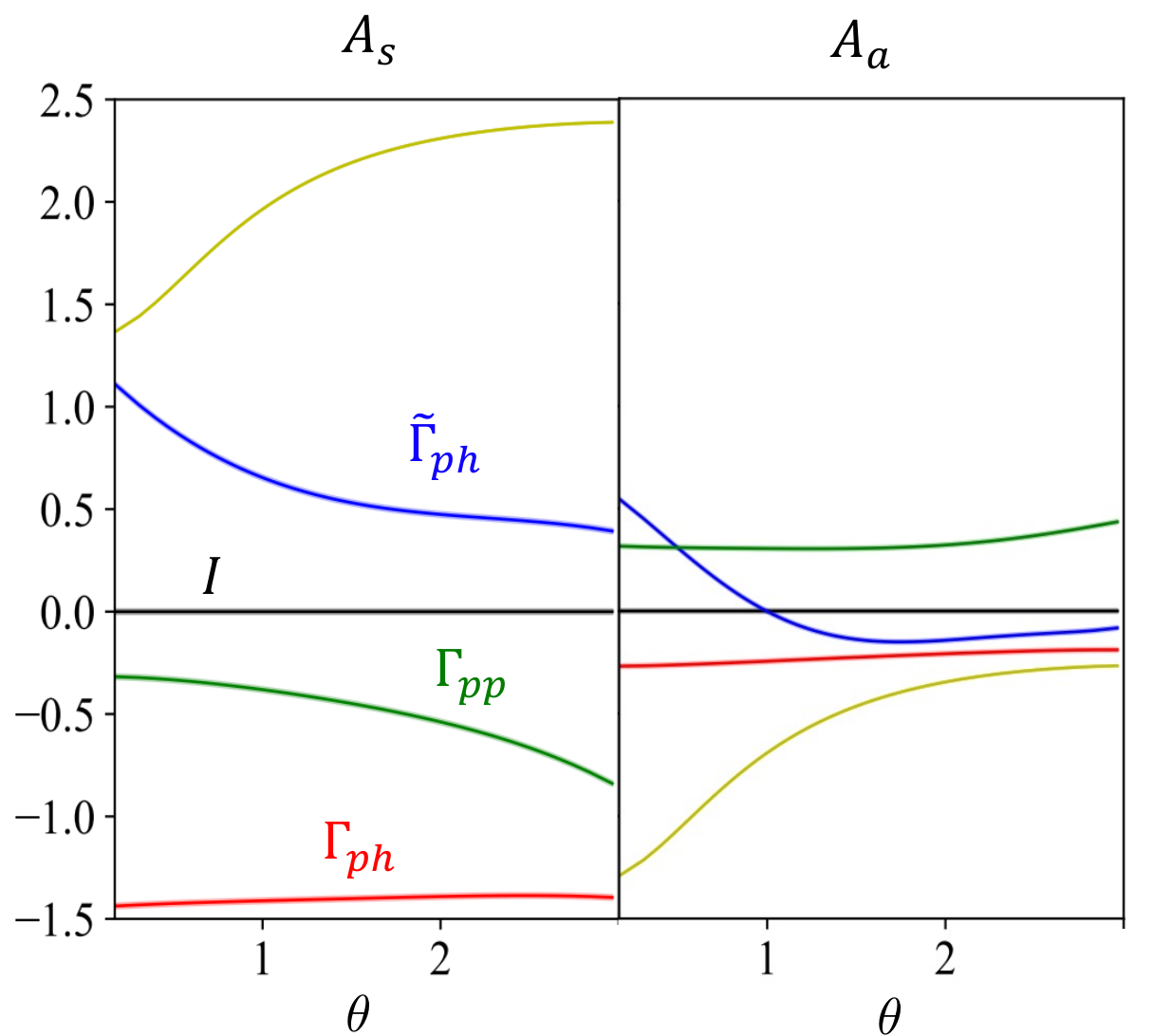}
\caption{\label{fig:ust} Decomposition of the angle-resolved scattering amplitude into different channels for the spin-symmetric component $A^s(\theta)$ (left panel) and the spin-asymmetric component $A^a(\theta)$ (right panel). The channels include the bare coupling (yellow), the particle-hole channel (red), the particle-hole-exchanged channel (blue), the particle-particle channel (green), and the fully irreducible channel (black). Significant sign cancellations between different channels explains a substantially reduced overall scattering amplitude compared to the bare coupling. Remarkably, the fully irreducible channel remains negligibly small even in the presence of strong coupling.}
\end{figure}

In stark contrast, the scattering amplitude exhibits a much smaller amplitude compared to the bare coupling. To elucidate the underlying mechanisms responsible for this suppression, we compute the contributions to the scattering amplitude from different channels, including the bare coupling (yellow), the particle-hole channel (red), the particle-hole-exchanged channel (blue), the particle-particle channel (green), and the fully irreducible channel (black). The results, presented in Fig. \ref{fig:ust}, reveal that the bare coupling is significantly cancelled out by the particle-hole contribution, indicating a strong screening effect. Moreover, we observe a substantial cancellation between the particle-hole-exchange channel and the particle-particle channel, further contributing to the small magnitude of the scattering amplitude. This extra sign cancellation contributes to the Fermi liquid stability by further reducing the quasiparticle interactions. These findings shed light on the intricate interplay between various many-body contributions and their role in renormalizing the effective interaction strength in strongly correlated systems. 

The renormalized perturbation theory in this work provides a systematic framework for studying nonlocal quasiparticle interactions in Fermi liquids. By systematically generating high-order diagrams and efficiently evaluating them using diagrammatic Monte Carlo techniques, we accurately calculate the 4-vertex function and its various momentum-frequency slices, capturing the essential physics while maintaining a well-behaved perturbative expansion. The emergent locality of the four-vertex function, the substantial enhancement of the Landau quasiparticle interaction, and the significant suppression of the scattering amplitude through channel cancellations provide valuable insights into the fundamental physics governing Fermi liquids.

\section{Conclusion}

In conclusion, our work explores a novel approach for investigating nonlocal effective interactions in quantum many-body systems. By selectively resumming the essential nonlocal contributions to the effective interaction vertex, we have developed a systematic tool for unraveling the complex physics of strongly correlated systems. This approach offers a significant advantage over traditional skeleton diagrammatic methods, which encounter the ``vertex problem" when attempting to resum the full interaction vertex. It opens up an alternative avenue for exploring the rich phenomena emerging from nonlocal effective interactions across various fields, from electronic structure~\cite{Giuliani_Vignale_2005} to ultracold atomic gases~\cite{shi2023} and nuclear matter~\cite{neutron_star_matter, nuclear_force}. We anticipate that the proposed renormalization scheme will provide valuable insights into the intricate interplay between nonlocality and strong correlations, deepening our understanding of quantum many-body systems.

The cornerstone of our approach is the introduction of a projection-based renormalization condition that selectively resums the most essential nonlocal contributions to the effective interaction vertex. This condition allows us to capture the crucial physics of interaction renormalization while avoiding the computational complexity associated with the full vertex function. Building upon this foundation, we have derived a renormalized perturbation theory that generates Feynman diagrams in powers of the renormalized nonlocal interaction.  To efficiently generate the vast space of renormalized diagrams, we have developed an algorithm utilizing a perturbative expansion of the parquet equations~\cite{hou2024feynman}. This approach enables the systematic construction of high-order diagrams by combining lower-order building blocks. To address the separate computational challenge of evaluating these diagrams, we have implemented a state-of-the-art DiagMC algorithm that employs a computational graph representation to compute the high-dimensional integrands associated with the diagrams.

The application of our renormalization scheme to a 3D Yukawa Fermi liquid has revealed significant insights into the role of nonlocal quasiparticle interactions and the importance of properly renormalizing them in quantum many-body systems. By deriving the precise quasiparticle interaction from the microscopic theory, we have demonstrated that the renormalized perturbation theory remains predictive even in the strongly correlated regime where the bare perturbation theory fails. Remarkably, our work uncovers significant cancellations between different channels contributing to the scattering amplitude. In addition to the cancellation caused by the particle-hole screening,  we observe a substantial cancellation between the particle-hole-exchange channel and
the particle-particle channel, providing a deeper understanding of how Fermi liquids maintain their stability and exhibit emergent weakly interacting behavior despite strong bare couplings.

Looking ahead, two exciting directions emerge from our work. First, our renormalization scheme provides a systematic framework for exploring nonlocal QFTs of a wide range of quantum many-body systems, including Fermi liquids, non-Fermi liquids, and beyond. Historically, the ``vertex problem" has posed a formidable challenge in performing high-order calculations and deriving precise predictions of low-energy properties from first principles in these theories. Combining the renormalization technique with existing proposals of effective field theories for Fermi liquids~\cite{Lee_FL_1} and non-Fermi liquids~\cite{Lee_FL_2} may enable high-order calculations of experimentally relevant observables. This could provide a rigorous test of the validity of these theories and shed light on the underlying mechanisms governing the behavior of strongly correlated fermionic systems. Furthermore, our renormalization scheme opens up new possibilities for constructing nonlocal QFTs to model systems that are difficult to capture using traditional local QFTs. A prime example is the electron liquid, where the long-range Coulomb interaction and the presence of a Fermi surface give rise to complex nonlocal correlations. By designing nonlocal QFTs specifically tailored to incorporate these correlations and employing our renormalization scheme to systematically compute high-order corrections, it may be possible to develop a more accurate and predictive theory of the conduction electrons in real materials.

Second, our renormalization scheme can help to solve the vertex problem in existing skeleton diagrammatic techniques. One prominent example is the fRG approach~\cite{fRG_review}, which has been widely used to study phase transitions and critical phenomena in quantum many-body systems. In some fRG implementations, the full vertex function is typically stored in memory using a discrete set of mesh points or basis functions, which inevitably introduces discretization errors. By defining the projection operator as the discretization protocol, one can establish a renormalization condition that gives rise to a renormalized perturbation theory. This theory provides a systematic framework for computing corrections to mitigate the discretization error. The renormalization condition ensures that the discretized vertex function captures the essential physics, while the corrections account for the discrepancies arising from the discretization. We anticipate that this technique will prove valuable for fRG and other skeleton methods that involve the manipulation of the full vertex function, enhancing their accuracy and reliability. 

In addition to the exciting applications of our renormalization scheme, there are several fundamental problems that warrant further investigation. From a theoretical perspective, it is crucial to explore how the projection renormalization condition can be designed to ensure that the resulting renormalized diagrammatic series respects conservation laws and Ward identities, generalizing the Baym-Kadanoff algorithm~\cite{Baym1,Baym2} to our partial renormalization scheme. Furthermore, the issue of wrong convergence, which plagues the skeleton diagrammatic series, may also affect our renormalized diagrammatic series and demands further investigation.  In this regard, the homotopic action approach, as suggested by recent findings~\cite{homotopic}, could potentially provide a more reliable mathematical structure for the diagrammatic expansion, ensuring convergence to physically correct results. Exploring the connections between our renormalization scheme and the homotopic action approach may yield valuable insights into the underlying mathematical structure and help address the problem of wrong convergence. Comparing our nonlocal renormalization scheme with the traditional local QFT approach reveals additional avenues for exploration, such as extending the Callan-Symanzik equation~\cite{peskin2018introduction} to nonlocal QFTs to provide new insights into the RG evolution of nonlocal interactions and the emergence of effective low-energy theories. Addressing these challenges will deepen our understanding of the partial renormalization scheme and shed light on the intricate interplay between nonlocality, conservation laws, and RG flows in QFTs.

In conclusion, our work establishes a versatile framework for investigating nonlocal effective interactions in quantum many-body systems. By providing a systematic approach to renormalize the essential nonlocal contributions to the effective interaction vertex, we have opened new avenues for unraveling the complex physics of strongly correlated systems. We anticipate that our renormalization scheme will find broad applications across various fields, from condensed matter physics to ultracold atomic gases and nuclear physics, paving the way for a deeper understanding of the fascinating phenomena associated with nonlocal effective interactions in quantum many-body systems.

{\it Acknowledgements -- } 
We are grateful to Gabriel Kotliar, Kristjan Haule, Jan von Delft, Fabian Kugler, Nikolay Prokof'ev, Boris Svistunov, Youjin Deng, Christopher Amey, Yashar Komijani, and John Sous for stimulating discussions, valuable input, and suggestions that greatly improved the readability of this manuscript. We also thank Alessandro Toschi, Piers Coleman and Romain Vasseur for enlightening conversations that helped shape our understanding of the subject. This work was supported by the National Natural Science Foundation of China under Grant No. 12047503 and the Simons Foundation through the Many Electron Collaboration. K.C. acknowledges additional support from the National Natural Science Foundation of China under Grant No. 12047503.
\bibliographystyle{apsrev4-1}
\bibliography{ref}

\end{document}